\documentclass[fleqn,preprintnumbers]{elsart}

\usepackage{graphicx}
\usepackage{dcolumn}
\usepackage{bm}
\usepackage{amssymb}
\usepackage{indentfirst}
\usepackage{epsfig}
\usepackage{epsf}
\usepackage{graphicx}

\usepackage{amssymb}

\usepackage{indentfirst}

\usepackage{epsfig}

\usepackage{epsf}

\usepackage{graphicx}

\def\Vec#1{{\bf #1}}

\newcommand{\kp}{\mathbf{k}_\perp}

\begin{document}

\begin{flushright}
USM-TH-177\\
\end{flushright}
\bigskip\bigskip

\begin{frontmatter}

\title{Flavor separation of the Boer-Mulders function from unpolarized $\pi^- p$ and $\pi^- D$ Drell-Yan processes}

\author[utfsm]{Zhun Lu},
\author[pku]{Bo-Qiang Ma\corauthref{cor}} \ead{mabq@phy.pku.edu.cn}
\corauth[cor]{Corresponding authors.},
\author[utfsm]{Ivan Schmidt\corauthref{cor}} \ead{ivan.schmidt@usm.cl}
\address[utfsm]{Departamento de F\'\i sica, Universidad T\'ecnica Federico
Santa Mar\'\i a, Casilla 110-V, Valpara\'\i so, Chile}
\address[pku]{School of Physics, Peking University, Beijing 100871, China}

\begin{abstract}
We show that measuring the $\cos 2\phi $ angular dependence in
unpolarized Drell-Yan processes with $\pi^-$ beams colliding on
proton and deuteron targets can determine the ratio of the
Boer-Mulders functions for $d$ and $u$ quarks inside the proton
$h_1^{\perp,d}/h_1^{\perp,u}$, which is still lack of theoretical
constraint. The comparison of the $\cos 2 \phi$ asymmetries
measured in unpolarized $\pi^- p$ and $\pi^- D$ Drell-Yan
processes, which are accessible at CERN by the COMPASS
collaboration, can help to discriminate whether $h_1^\perp$
effects or QCD vacuum effects are preferred by data.

\end{abstract}


\end{frontmatter}

\vspace{1cm}


 The transverse spin phenomena appearing in high energy
scattering processes~\cite{bdr} is among the most interesting issues
of spin physics. Substantial single-spin asymmetries (SSA) in
semi-inclusive deeply inelastic scattering
(SIDIS)~\cite{smc,Airapetian:2004tw,compass,hermes05}, with one
incoming nucleon transversely polarized, have been measured by
several experiments. The interpretation of these asymmetries
provides new insights into QCD and nucleon
structure~\cite{sivers,collins93,anselmino95,bm,bhs02,collins02,belitsky}.
One of the mechanisms which can account for such SSA is the Sivers
effect~\cite{sivers}, related to a spin and $\Vec k_T$-dependent
function~\cite{levelt,kotzinian,mulders} named as Sivers function
$f_{1T}^\perp(x,\Vec k_T^2)$. It arises from the correlation between
the nucleon transverse spin and quark transverse momentum. Despite
its naively $T$-odd structure, $f_{1T}^\perp$ has been shown to be
non-zero~\cite{bhs02} due to its special gauge-link
structure~\cite{collins02,belitsky}. In the case of unpolarized
collisions, large $\cos 2 \phi$ asymmetries have been observed
~\cite{na10,conway} in $\pi^- N$ Drell-Yan dilepton production
processes. A parton interpretation of this asymmetry has been
proposed in~\cite{boer} in terms of another leading twist $T$-odd
distribution function, the Boer-Mulders function $h_{1}^\perp(x,\Vec
k_T^2)$~\cite{bm}, which has the same QCD origin~\cite{collins02} as
the Sivers function. It describes the transverse polarization
distribution of the quark inside the unpolarized hadron due to the
correlation between the quark transverse spin and transverse
momentum. The model calculations of the Boer-Mulders function have
been performed in Refs.~\cite{gg02,bbh03,yuan,bsy04} for the proton
and in Ref.~\cite{lm04} for the pion, based on the gauge-link
structure of the distribution. The resulting asymmetries have been
estimated for unpolarized $p\bar{p}$~\cite{bbh03,gg05,blm06} and
$\pi^- N$ ~\cite{lm05} Drell-Yan processes. The role of Boer-Mulders
functions in unpolarized SIDIS process has also been
discussed~\cite{gg03,blm05}. One of the significant features of the
Boer-Mulders function relies on that fact that the spin structure of
hadrons can be also studied in physical processes without invoking
beam or target polarization.

The large anomalous Drell-Yan asymmetry implies a significant
non-zero size of the Boer-Mulders function. However its extraction
from experimental data is relatively difficult comparing to that
of the Sivers function. In the latter case the Sivers function is
convoluted with ordinary distribution/fragmentation functions,
which are well-known. Several sets of the proton Sivers functions
for both $u$ and $d$ quark have been
extracted~\cite{anselmino05c,vy05,efremov,anselmino05b} recently
from HERMES~\cite{Airapetian:2004tw,hermes05} and
COMPASS~\cite{compass} data on single-transverse spin asymmetries
in SIDIS. 
In the process in which the Boer-Mulders function contributes, it
is always convoluted with itself or other chiral-odd functions,
such as the transversity or the Collins fragmentation
function~\cite{collins93}, which are also not well known at
present. Different models or theoretical
considerations~\cite{yuan,bsy04,pobylista} predict very different
flavor dependence of $h_1^\perp$, and even the relative sign
between $h_1^{\perp,u}$ and $h_1^{\perp,d}$ is not known. In this
paper we suggest to use unpolarized $\pi^- p$ and $\pi^- D$
 Drell-Yan processes to access the flavor
dependence of $h_1^\perp$, especially the ratio
$h_1^{\perp,d}/h_1^{\perp,u}$. The hadron program by the COMPASS
collaboration will start in 2007 at CERN, in which a $\pi^-$ beam
colliding with both proton and deuteron targets are going to be
available. We show that by comparing the $\cos 2 \phi$ angular
dependent part of the cross sections of unpolarized $\pi^- p$ and
$\pi^- D$ Drell-Yan processes, one can determine the ratio
$h_1^{\perp,d}/h_1^{\perp,u}$. By comparing the $\cos 2 \phi$
asymmetries in these processes, one can also discriminate whether
$h_1^\perp$ effects or QCD vacuum effects~\cite{bnm93} are preferred
by data. The later effects can also explain the observed asymmetry
in a quite different way. The idea is that the non-perturbative
vacuum structure in QCD can lead to fluctuating chromomagnetic
vacuum fields. When the $q$ and $\bar{q}$ travel through these
fields the spin orientations of the quark pair might be
correlated~\cite{nr84} before they annihilate to $\gamma^*$. In
Ref.~\cite{bnm93} a factorization violating spin correlation is
proposed and used to fit the data of the NA10 experiments. According
to the color interaction nature of these vacuum effects, the
``correlation strengths" for different quark flavors are the same.
Therefore vacuum effects imply no flavor dependence of the
asymmetries in different processes, which is different from
$h_1^\perp$ effects.

The general form of the angular differential cross section for
unpolarized Drell-Yan process is
\begin{eqnarray}
\frac{1}{\sigma}\frac{d\sigma}{d\Omega}&=&\frac{3}{4\pi}\frac{1}{\lambda+3}
\left
(1+\lambda\textmd{cos}^2\theta+\mu\textmd{sin}2\theta\textmd{cos}\phi
+\frac{\nu}{2}\textmd{sin}^2\theta\textmd{cos}2\phi\right ).
\end{eqnarray}

The $\cos2\phi$ angular dependence of the lepton pair can be
analyzed in the Collins-Soper frame, in which the unpolarized
differential cross section is expressed as
\begin{eqnarray}
&&\hspace{-1cm}\frac{d\sigma(h_Ah_B\rightarrow l\bar{l}X)}{d\Omega
dx_1dx_2d^2\mathbf{q}_\perp}=
\frac{\alpha^2}{3Q^2}\sum_ae_a^2\Bigg{\{}
A(y)\mathcal{F}[f_{1a/A}(x_1,\mathbf{p}_\perp^2)f_{1\bar{a}/B}(x_2,\mathbf{k}_\perp^2)] +B(y)\textmd{cos}2\phi\nonumber\\
&&\hspace{-1cm}\times\mathcal{F}\left [(2\hat{\mathbf{h}}\cdot
\mathbf{p}_\perp\hat{\mathbf{h}}\cdot \mathbf{k}_\perp
-\mathbf{p}_\perp\cdot
\kp)\frac{h_{1a/A}^\perp(x_1,\mathbf{p}_\perp^2)
h_{1\bar{a}/B}^\perp(x_2,\mathbf{k}_\perp^2)}{M_AM_B}\right
]\Bigg{\}}+(a \leftrightarrow \bar{a}), \label{cs}
\end{eqnarray}
where the notation
\begin{eqnarray}
\mathcal{F}[\cdots]&=&\int d^2\mathbf{p}_\perp
d^2\kp\delta^2(\mathbf{p}_\perp+\kp-\mathbf{q}_\perp)\times\{\cdots\}
\end{eqnarray}
shows the convolution of transverse momenta, and
\begin{eqnarray}
A(y)&=&\left ( \frac{1}{2}-y+y^2 \right )=\frac{1}{4}(1+\cos^2\theta),\\
B(y)&=&y(1-y)=\frac{1}{4}\sin^2\theta
\end{eqnarray}
in the c.s. frame of the lepton pair. From Eq.~(\ref{cs}) we can
write down the expression for the $\cos2\phi$ asymmetry
coefficient ($\lambda=1$, $\mu=0$)
\begin{eqnarray}
\nu=\frac{2\mathcal{F}\left [(2\hat{\mathbf{h}}\cdot
\mathbf{p}_\perp\hat{\mathbf{h}}\cdot \mathbf{k}_\perp
-\mathbf{p}_\perp\cdot \kp){h_{1a/A}^\perp
h_{1\bar{a}/B}^\perp}\right
]}{{M_AM_B}\mathcal{F}[f_{1a/A}f_{1\bar{a}/B}]}.
\end{eqnarray}

The azimuthal angel dependent part can be picked up by a weighting
procedure~\cite{km97}:
\begin{eqnarray}
\left \langle \frac{q_T^2 \cos 2 \phi}{4M_A M_B} \right \rangle &
= & \int d\phi d^2q_T \frac{d\sigma(h_Ah_B\rightarrow
l\bar{l}X)}{d\Omega dx_1dx_2d^2\mathbf{q}_\perp} \cdot \frac{q_T^2
\cos 2 \phi}{4M_A M_B}\label{weight}\\
&=&\frac{\alpha^2}{3Q^2}B(y)\sum_ae_a^2h_{1a/A}^{\perp(1)}(x_1)h_{1\bar{a}/B}^{\perp(1)}(x_2)+(a
\leftrightarrow \bar{a}),
\end{eqnarray}
in which
\begin{equation}
h_{1a/H}^{\perp(1)}(x)=\int d^2k_\perp \frac{k^2_\perp}{2M^2_H}
h_{1a/H}^\perp(x,k^2_\perp)
\end{equation}
is the first $k^2_\perp$-moment of $h_{1a/H}^\perp(x,k^2_\perp)$.

\begin{figure}

\begin{center}
\scalebox{0.9}{\includegraphics{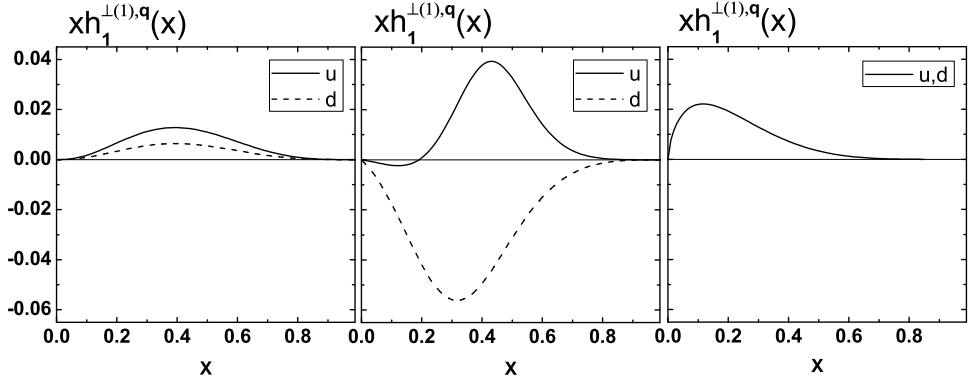}} \caption{\small
$xh_1^{\perp(1),u}(x)$ and $xh_1^{\perp(1),d}(x)$ from different
models or theoretical considerations: (a) MIT bag model, (b)
spectator model with axial-vector diquark, and (c) large-$N_c$
limit.} \label{bmud}
\end{center}

\end{figure}

Because of the simple structure of the pion compared to that of
the nucleon, it is advantageous to use a pion beam to unravel the
quark structure of the nucleon in the Drell-Yan process. In
leading twist there are only two distribution functions for the
pion, the usual momentum distribution function and the
Boer-Mulders function. In Refs.~\cite{lm04,lm05} calculations
based on a quark spectator antiquark model showed that the
Boer-Mulders functions of the two valence quarks inside the pion
(such as $\bar{u}$ and $d$ inside $\pi^-$) are equal. Actually
this relation can be obtained from model independent principles:
charge conjugation symmetry and isospin symmetry. In the following
we will use $h_{1\pi}^{\perp}$ and $\bar{h}_{1\pi}^{\perp}$ to
denote the Boer-Mulders functions of the valence and sea quarks
inside the pion respectively, and $h_{1}^{\perp,{q(\bar{q})}}$ for
the Boer-Mulders functions of the quark flavor $q(\bar{q})$ inside
the proton. Thus the weighted cross section of the unpolarized
$\pi^- p$ Drell-Yan process is ($W=Q_T^2 \cos 2 \phi/4M_A M_B$)
\begin{eqnarray}
\left \langle W\right \rangle_{\pi^-
p}(x_1,x_2)&=&\frac{\alpha^2}{3Q^2}B(y)\left
\{h_{1\pi}^{\perp(1)}(x_1)\left [e_u^2h_{1}^{\perp(1),u}(x_2)
+e_d^2h_{1}^{\perp(1),\bar{d}}(x_2))\right ]\right.\nonumber\\
&&\left.+\bar{h}_{1\pi}^{\perp(1)}(x_1)\left
[e_u^2h_{1}^{\perp(1),\bar{u}}(x_2)
+e_d^2h_{1}^{\perp(1),d}(x_2))\right ]\right \}.
\end{eqnarray}

We are interested in the region were $x_{1/2}$ are not so small,
where the sea quark contribution to the $\cos2\phi$ angular
dependent cross section is expected to be small. Then we can use
the approximation
\begin{eqnarray}
\left \langle W\right \rangle_{\pi^-
p}(x_1,x_2)\approx\frac{4\alpha^2}{3Q^2}B(y)e_u^2h_{1\pi}^{\perp(1)}(x_1)h_{1}^{\perp(1),u}(x_2).
\end{eqnarray}

The cross section for the $\pi^- D$ Drell-Yan process can be
approximated as $\sigma_{\pi^- D} = \sigma_{\pi^- p}+\sigma_{\pi^-
n}$, thus
\begin{eqnarray}
\langle W \rangle_{\pi^- D}(x_1,x_2)=\langle W \rangle_{\pi^-
p}(x_1,x_2)+\langle W \rangle_{\pi^- n}(x_1,x_2).
\end{eqnarray}
The Boer-Mulders functions of quarks inside the neutron can be
related to those in the proton: $h_{1u/n}^{\perp}=h_1^{\perp,d}$,
$h_{1d/n}^{\perp}=h_1^{\perp,u}$. Therefore
\begin{eqnarray}
 \left \langle W\right \rangle_{\pi^-
n}(x_1,x_2)&=&\frac{\alpha^2}{3Q^2}B(y)\left
\{h_{1\pi}^{\perp(1)}(x_1)\left [e_u^2h_{1}^{\perp(1),d}(x_2)
+e_d^2h_{1}^{\perp(1),\bar{u}}(x_2))\right ]\right.\nonumber\\
&&\left.+\bar{h}_{1\pi}^{\perp(1)}(x_1)\left
[e_u^2h_{1}^{\perp(1),\bar{d}}(x_2)
+e_d^2h_{1}^{\perp(1),u}(x_2))\right ]\right \}.
\end{eqnarray}
Ignoring the sea quark contribution, we have
\begin{eqnarray}
\left \langle W\right \rangle_{\pi^-
D}(x_1,x_2)\approx\frac{\alpha^2}{3Q^2}B(y)e_u^2h_{1\pi}^{\perp(1)}(x_1)
(h_{1}^{\perp(1),u}(x_2)+h_{1}^{\perp(1),d}(x_2)).
\end{eqnarray}
Then the ratio of the weighted cross sections for unpolarized
$\pi^- p$ and $\pi^- D$ Drell-Yan processes is
\begin{eqnarray}
\frac{\langle W \rangle_{\pi^- D}(x_1,x_2)}{2 \langle W
\rangle_{\pi^- p}(x_1,x_2)}=\frac{1}{2} \left (
1+\frac{h_{1}^{\perp(1),d}(x_2)}{h_{1}^{\perp(1),u}(x_2)} \right
).\label{Doverp}
\end{eqnarray}

The Drell-Yan process at COMPASS can explore the kinematics region
$s=400\, \textrm{GeV}^2$ and $Q^2=20 \, \textrm{GeV}^2$, where
$x_1 x_2=Q^2/s=0.05$. Therefore the Drell-Yan experiments at
COMPASS can access the quark structure in the valence regime $0.1
\leq x_{1/2} \leq 0.5$, where Eq.~(\ref{Doverp}) is a good
approximation. Then the ratio of the Boer-Mulders functions of $d$
and $u$ quarks is
\begin{eqnarray}
\frac{h_{1}^{\perp(1),d}(x_2)}{h_{1}^{\perp(1),u}(x_2)}=\frac{\langle
W \rangle_{\pi^- D}(x_1,x_2)}{\langle W \rangle_{\pi^-
p}(x_1,x_2)}-1.\label{doveru}
\end{eqnarray}

Another possibility is to use a $\pi^+$ beam to collide on the
proton, which process is sensitive to the $d$ quark distribution
of the proton:
\begin{eqnarray}
\frac{\langle W \rangle_{\pi^+ p}(x_1,x_2)}{\langle W
\rangle_{\pi^-
p}(x_1,x_2)}=\frac{h_{1}^{\perp(1),d}(x_2)}{4h_{1}^{\perp(1),u}(x_2)}.\label{plovermi}
\end{eqnarray}
This relation can serve as cross check for Eq.~(\ref{doveru}).

knowledge of the Boer-Mulders functions is very poor, including
its size and shape, and the relative size between $h_1^{\perp,u}$
and $h_1^{\perp,d}$. In the last case different models or
theoretical considerations predict quite different values for the
ratio $h_1^{\perp,d}/h_1^{\perp,u}$. A calculation based on the
MIT bag model gave~\cite{yuan}
$h_1^{\perp,d}=\frac{1}{2}h_1^{\perp,u}$. The spectator model with
axial-vector diquarks predicted~\cite{bsy04} that $h_1^{\perp,d}$
has a opposite sign and a rather larger size compared with
$h_1^{\perp,u}$, while large-$N_c$ showed that
$h_1^{\perp,d}=h_1^{\perp,u}$~\cite{pobylista} modulo $1/N_c$
corrections. In Fig.~\ref{bmud} we show the curves for
$h_1^{\perp(1),u}(x)$ and $h_1^{\perp(1),d}(x)$ from these models
or theoretical considerations. In the case of large-$N_c$ since
there is no detail calculation on the size of $h_1^\perp$, we just
naively assume $h_1^{\perp(1),u}=f_{1T}^{\perp(1),u}$ (these two
function are closely related and expected to have similar size),
and for $f_{1T}^{\perp(1),u}$ we adopt the
parametrization~\cite{efremov} extracted from HERMES experimental
data based on large-$N_c$. In the case of MIT bag model the ratio
in Eq.~(\ref{Doverp}) is $0.75$ and in large-$N_c$ consideration
is 1, while in axial-vector diquark model the ratio is mostly
negative. These predictions can be tested by measuring the ratio
$\langle W \rangle_{\pi^- D}/(2\langle W \rangle_{\pi^- p})$. Vice
versa, according to Eq.~(\ref{doveru}), from the measured ratio
$\langle W \rangle_{\pi^- D}/(2\langle W \rangle_{\pi^- p})$ the
size of $h_1^{\perp,d}/h_1^{\perp,u}$ can be determined.

The analysis can be easily extended to the single (transverse)
polarized Drell-Yan processes $\pi p^\uparrow(D^\uparrow)
\rightarrow \mu^+\mu^- X$ at COMPASS. The measurement of the
single-spin asymmetries in this process is also important for the
more precise extraction of the Sivers function (through
$f_{1\pi}\otimes f_{1T}^\perp$), and to test the QCD prediction
$f_{1T}^\perp|_{\small \textrm{SIDIS}}=-f_{1T}^\perp|_{\small
\textrm{DY}}$~\cite{collins02,efremov2}, and can provide
alternative access to transversity (through $h_{1\pi}^\perp\otimes
h_1$)~\cite{boer}. In single polarized Drell-Yan processes the
relations similar to Eqs.~(\ref{Doverp}) and (\ref{plovermi}) can
be deduced for the Sivers function and for transversity.

\begin{figure}

\begin{center}
\scalebox{0.82}{\includegraphics{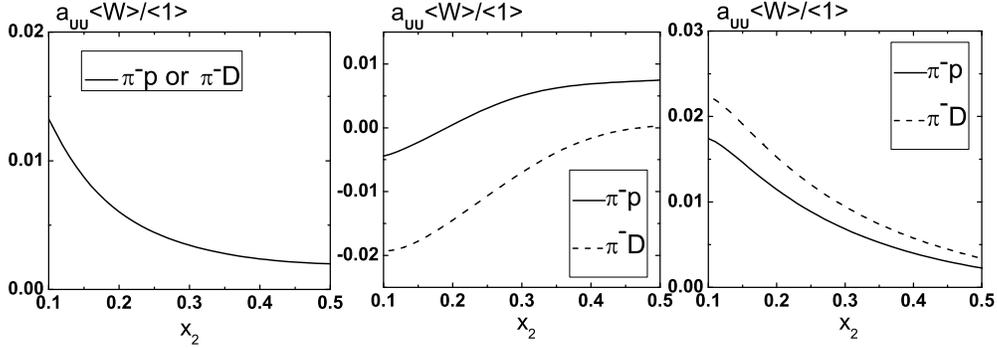}} \caption{\small The
weighted $\cos 2 \phi$ asymmetries in unpolarized $\pi ^- p$
(solid line) and $\pi^- D$ (dashed line) Drell-Yan processes. (a)
The MIT Bag model result. The two asymmetries are equal since
$f_1^d/f_1^u=1/2$ in the lowest bag mode in which case is used in
calculation. (b) The axial-vector diquark model result. (c) The
result from large-$N_c$.} \label{weiasy}
\end{center}

\end{figure}

It is also interesting to compare the $\cos 2 \phi$ asymmetries in
unpolarized $\pi^- p$ and $\pi^- D$ processes. The anomalous
Drell-Yan asymmetry might be produced by $h_1^\perp$, which is
labelled as hadronic effect. Nevertheless, there is another effect
that could also account for this asymmetry, the so called QCD
vacuum effect~\cite{bnm93}. A significant difference between these
two effects~\cite{bbnu} is that the first one is flavor dependent,
while the latter is flavor blind. Thus comparing the asymmetries
in different processes can discriminate which effect is really
responsible for the asymmetry. In the case of the hadronic effect,
the weighted $\cos 2\phi$ asymmetries of these two processes can
be calculated from following expressions:
\begin{eqnarray}
\hat{\nu}_p&=& a_{UU}\frac{\langle W \rangle_{\pi^- p}}{\langle\,
1 \rangle_{\pi^- p}} =
\frac{h_{1\pi}^{\perp(1)}(x_1)h_{1}^{\perp(1),u}(x_2)}{f_{1\pi}(x_1)f_1^u(x_2)},\label{asyp}\\
\hat{\nu }_D&=& a_{UU}\frac{\langle W \rangle_{\pi^- D}}{\langle
\, 1\rangle_{\pi^- D}}=
\frac{}{}\frac{h_{1\pi}^{\perp(1)}(x_1)(h_{1}^{\perp(1),u}(x_2)+
h_{1}^{\perp(1),d}(x_2))}{f_{1\pi}(x_1)(f_1^u(x_2)+f_1^d(x_2))},\label{asyd}
\end{eqnarray}
where $a_{UU}=(1+\cos^2\theta)/\sin^2\theta$, and $\langle 1
\rangle$ is calculated from Eq.~(\ref{weight}) when $W$ is
replaced by $1$. Again we neglect the sea contribution here. The
ratio of the asymmetries in these two processes is then
\begin{eqnarray}
\frac{\hat{\nu}_D}{\hat{\nu
}_p}=\frac{1+\frac{h_1^{\perp(1),d}(x_2)}{h_1^{\perp(1),u}(x_2)}}{1+\frac{f_1^d(x_2)}{f_1^u(x_2)}}.
\end{eqnarray}
The QCD vacuum effect will yield $\hat{\nu}_p=\hat{\nu}_D$. In
Fig.~\ref{weiasy} we show the asymmetries calculated from
Eq.~(\ref{asyp}) and (\ref{asyd}) using the three options of
Boer-Mulders functions given in Fig.~\ref{bmud}. For the
unpolarized distributions, which appear in the denominators in
Eq.~(\ref{asyp}) and (\ref{asyd}) we adopt well known model
results. We do not aim at a precise estimate of the asymmetries,
only to a rough comparison among them. In the case that
$h_1^{\perp(1),u}/f_1^u$ is different from
$h_1^{\perp(1),d}/f_1^d$ (equivalently
$h_1^{\perp(1),d}/h_1^{\perp(1),u}$ different from $f_1^d/f_1^u$),
different asymmetries will be observed in these two processes, as
shown in Fig.~\ref{weiasy}b and Fig.~\ref{weiasy}c. Therefore the
difference appearing in the $\cos 2 \phi$ asymmetries in
unpolarized $\pi^- p$ and $\pi^- D$ Drell-Yan processes can
provide evidence whether hadronic effects or QCD vacuum effects
are preferred by data.

In summary, we have showed that by comparing the $\cos 2 \phi$
angular dependent part of the cross sections of the unpolarized
$\pi^- p$ and $\pi^- D$ Drell-Yan processes, which will be
accessible in future COMPASS experiments, one can determine the
ratio $h_1^{\perp,d}/h_1^{\perp,u}$ in the valence region, which
still lacks theoretical input. By checking the difference of the
$\cos 2 \phi$ asymmetries in these two processes, one can also
provide a discrimination whether hadronic effects or QCD vacuum
effects are preferred by data. The investigation will lead to
better understanding about the role of the quark transverse motion
and transverse spin in hadronic reactions.

{\bf Acknowledgements.} This work is partially supported by
National Natural Science Foundation of China (Nos.~10421503,
10575003, 10505001, 10528510), by the Key Grant Project of Chinese
Ministry of Education (No.~305001), by the Research Fund for the
Doctoral Program of Higher Education (China), by Fondecyt (Chile)
under Project No.~3050047 and No.~1030355.


\begin{thebibliography}{99}

\bibitem{bdr}

For a review on tranverse polarization phenomena, see V.~Barone,
A.~Drago, P.G.~Ratcliffe, Phys. Rep. 359 1 (2002).

\bibitem{smc} A. Bravar et al., SMC Collaboration, Nucl. Phys. A 666 (2000) 314.

\bibitem{Airapetian:2004tw}

A.~Airapetian et al., HERMES Collaboration, Phys. Rev. Lett. 94
(2005) 012002.



\bibitem{compass}

V.Yu.~Alexakhin et al., COMPASS Collaboration, Phys. Rev. Lett. 94
(2005) 202002.

\bibitem{hermes05} M. Diefenthaler, HERMES Collaboration, in Proceedings of DIS 2005, Madison, Wisconsin (USA), hep-ex/0507013.




\bibitem{sivers} D. Sivers, Phys. Rev. D 41 (1990) 83;\\
D. Sivers, Phys. Rev. D 43 (1991) 261.


\bibitem{collins93} J.C.~Collins, Nucl. Phys. B 396 (1993) 161.

\bibitem{anselmino95}

M. Anselmino, M. Boglione, F. Murgia, Phys. Lett. B 362
164 (1995). \\
 M. Anselmino, F. Murgia, Phys. Lett. B 442, 470
(1998).


\bibitem{bm}

D.~Boer, P.J.~Mulders, Phys. Rev. D 57 (1998) 5780.

\bibitem{bhs02} S.J.~Brodsky, D.S.~Hwang,
 I.~Schmidt, Phys. Lett. B 530 (2002) 99;\\
S.J.~Brodsky, D.S.~Hwang, I.~Schmidt, Nucl. Phys. B 642 (2002)
344.

\bibitem{collins02} J.C. Collins,  Phys. Lett. B 536 (2002) 43.



\bibitem{belitsky}
X.~Ji and F.~Yuan, Phys. Lett. B 543
(2002) 66;\\
A.V.~Belitsky, X.~Ji, F.~Yuan, Nucl. Phys. B 656 (2003) 165; \\
D. Boer, P.J. Mulders, F. Pijlman, Nucl. Phys. B 667 (2003) 201.


\bibitem{levelt}

J.~Levelt and P.J.~Mulders, Phys. Rev. D 49 (1994) 96.



\bibitem{kotzinian}

A.~Kotzinian, Nucl. Phys. B 441 (1995) 234.



\bibitem{mulders}

P.J.~Mulders and R.D.~Tangerman, Nucl. Phys. B 461 (1996)



\bibitem{na10} S.~Falciano et al. NA10 Collaboration,
Z. Phys. C 31 (1986) 513;\\
M.~Guanziroli et al. NA10 Collaboration, Z. Phys. C 37 (1988) 545.

\bibitem{conway} J.S.~Conway et al.
Phys. Rev. D 39 (1989) 92.

\bibitem{boer}

D.~Boer, Phys. Rev. D 60 (1999) 014012.




\bibitem{gg02}  G.R. Goldstein, L. Gamberg, Talk given at 31st International
Conference on High Energy Physics (ICHEP 2002), Amsterdam, The
Netherlands, 24-31 July 2002, hep-ph/0209085.

\bibitem{bbh03} D.~Boer, S.J.~Brodsky, D.S.~Hwang, Phys.
Rev. D 67 (2003) 054003.







\bibitem{yuan}
F. Yuan, Phys. Lett. B 575 (2003) 45.

\bibitem{bsy04} A.~Bacchetta, A.~Sch\"{a}fer,
 J.-J.~Yang, Phys. Lett. B 578 (2004) 109.





\bibitem{lm04}
Z.~Lu, B.-Q.~Ma, Phys. Rev. D 70 (2004) 094044.

\bibitem{gg05} G. R. Goldstein, L. Gamberg, hep-ph/0506127.

\bibitem{blm06} V. Barone, Z.~Lu, B.-Q.~Ma, hep-ph/0612350, to appear in European Physical Journal C.

\bibitem{lm05}
Z.~Lu, B.-Q.~Ma, Phys. Lett. B 615 (2005) 200.


\bibitem{gg03} L.P.~Gamberg, G.R.~Goldstein, K.A.~Oganessyan, Phys. Rev. D 67 (2003) 071504;\\
 L.P.~Gamberg, G.R.~Goldstein
, K.A.~Oganessyan, Phys. Rev. D 68 (2003) 051501(R).




\bibitem{blm05} V. Barone, Z. Lu, B.-Q. Ma, Phys. Lett. B 632
(2006) 277.

\bibitem{anselmino05c} M. Anselmino, M. Boglione, U. D'lesio, A. Kotzinian, F. Murgia,
 A. Prokudin, Phys. Rev. D 72 (2005) 094007;\\
 M. Anselmino, M. Boglione, U. D'lesio, A. Kotzinian, F. Murgia,
 A. Prokudin, Phys. Rev. D 72 (2005) 099903, Erratum.

\bibitem{vy05} W. Vogelsang , F. Yuan, Phys. Rev. D 72 (2005) 054028.


\bibitem{efremov}
J.C. Collins, A.V. Efremov, K. Goeke, S. Menzel, A. Metz, P.
Schweitzer,  Phys. Rev. D73 (2006) 014021;\\
J.C. Collins, A. V. Efremov, K. Goeke, M. Grosse Perdekamp, S.
Menzel, B. Meredith, A. Metz, P. Schweitzer, hep-ph/0510342.



\bibitem{anselmino05b} For a
comparison to this extraction, see M. Anselmino et al.,
hep-ph/0511017.

\bibitem{pobylista} P.V. Pobylitsa, hep-ph/0301236.

\bibitem{bnm93} A. Brandenburg, O. Nachtmann, E. Mirkes, Z. Phys. C 60 (1993) 697.

\bibitem{nr84} O. Nachtmann, A. Reiter, Z. Phys. C 24 (1984) 283.

\bibitem{km97}

A.~Kotzinian, P.J. Mulders, Phys. Letts. B 406 (1997) 373.

\bibitem{efremov2} A.V. Efremov, K. Goeke, S. Menzel, A. Metz, P. Schweitzer,
Phys. Lett. B 612 (2005) 233.

\bibitem{bbnu} D. Boer, A. Brandenburg, O. Nachtmann,  Utermann, Eur.
Phys. J. C 40 (05) 55.











\end{thebibliography}
\end{document}